# Global permanent deformations triggered by the Sumatra earthquake

E. Boschi, E. Casarotti, R. Devoti, D. Melini, A. Piersanti, G. Pietrantonio, F. Riguzzi

Istituto Nazionale di Geofisica e Vulcanologia, via di Vigna Murata 605, Roma (Italy), e-mail: piersanti@ingv.it

The devastating megathrust earthquake of December 26 2004 off the west coast of northern Sumatra has been probably the largest since the 1960 Chile event. Its moment magnitude has been estimated to be 9.0 (corresponding to a seismic moment release of $4 \times 10^{22}$ N m) using surface wave data but some researchers suggest that about 2/3 of the elastic energy has been emitted aseismically exciting only extremely low frequency normal modes (1). This event was energetic enough to have detectable effects on Earth rotational parameters. Very preliminary calculations, taking into account only the high frequency energy emission and consequently underestimating global effects show that the Sumatra earthquake should have produced a pole shift large enough to be identified in the observed data series, a small change in the length of the day and a change in the oblateness of the Earth (2).

Though there have been, in the last years, several numerical results indicating that the permanent deformation field associated with giant earthquakes could be detectable on extremely large scale (comparable with plate dimension), until now, extremely far field post-earthquake deformations have never been detected except for a single controversial observation associated with the Alaska 1964 event (3, 4). Now, the Sumatra event represents a unique candidate to test this hypothesis.

With this aim, the world-wide network of permanent IGS sites have been examined using the weekly averaged time series available at the SOPAC data centre (ftp:\\garner.ucsd.edu). After constraining the weekly GPS solutions to the ITRF2000 frame, the time series have been subject to a hypothesis test (t-test) to detect the presence of coseismic offsets (Fig. S1, Tab. S1). Significant displacements are detected for GPS sites covering a vast region around the epicenter location in an area exceeding $10^7$ km$^2$: All the previous instrumental evidence of long distance seismic residual permanent deformations covered an area below $10^5$ km$^2$ (5).

We have modeled the residual permanent deformation associated with the earthquake using a spherical model of global co- and postseismic deformation (4). The results agree with the observed deformation quite well in several sites while the fit is poor especially in Indian sites (Fig. 1). In general, the sites where the fit is not good display a greater amount of observed displacement with respect to the computed one. This could confirm the hypothesis of a considerable amount of deformation energy released aseismically (1). Here we present a preliminary attempt to include some aseismic moment release on three additional planes with respect to the one originating the seismic rupture (6) the first one represents the extension of the fault plane along the slab, the remaining two correspond to the aftershock area (1). The inclusion of aseismic effects led to an improvement in the quality of the fit for most of the sites even if for few stations it remains fairly poor and await for more sophisticated source inversion modeling. Although very much numerical modeling effort in the future is needed to precisely describe the residual permanent deformation field caused by this giant event and to assess the role played by aseismic energy release and long term postseismic displacements, the observations and numerical modeling already available allow us to affirm that the Sumatra earthquake excited a permanent detectable deformation field on such a great spatial scale that its effects can be considered as almost global.

**References and notes**

1. S. Stein, E. Okal, Power of tsunami earthquake heavily underestimated, *New Scientist*, http://www.newscientist.com/article.ns?id=dn6991 (2005).

2. B. F. Chao, Did the 26 December 2004 Sumatra, Indonesia, Earthquake Disrupt the Earth's Rotation as the Mass Media Have Said?, *Eos* **86**, 01 (2005)

3. F. Press, Displacements, strains, and tilts at teleseismic distances, *J. Geophys. Res*. **70**, 2395 (1965).

4. A. Piersanti, G. Spada, R. Sabadini, Global postseismic rebound of a viscoelastic Earth: Theory for ¢nite faults and application to the 1964 Alaska earthquake, *J. Geophys. Res*. **102**, 477 (1997).

5. A. Piersanti, C. Nostro, F. Riguzzi, Active displacement field in the Suez-Sinai area: the role of postseismic deformation, *Earth Plan. Sci. Lett.* **193**, 13 (2001).

6. U.S. Geological Survey, *Magnitude 9.0 OFF THE WEST COAST OF NORTHERN SUMATRA Sunday, December 26, 2004 at 00:58:53 UTC* (1998, http://neic.usgs.gov/neis/eq_depot/2004/eq_041226/neic_slav_ff.html)

7. We thank the SOPAC team for providing the GPS weekly combined solutions publicly.

**Supporting Online Material**

Fig. S1

Tab. S1

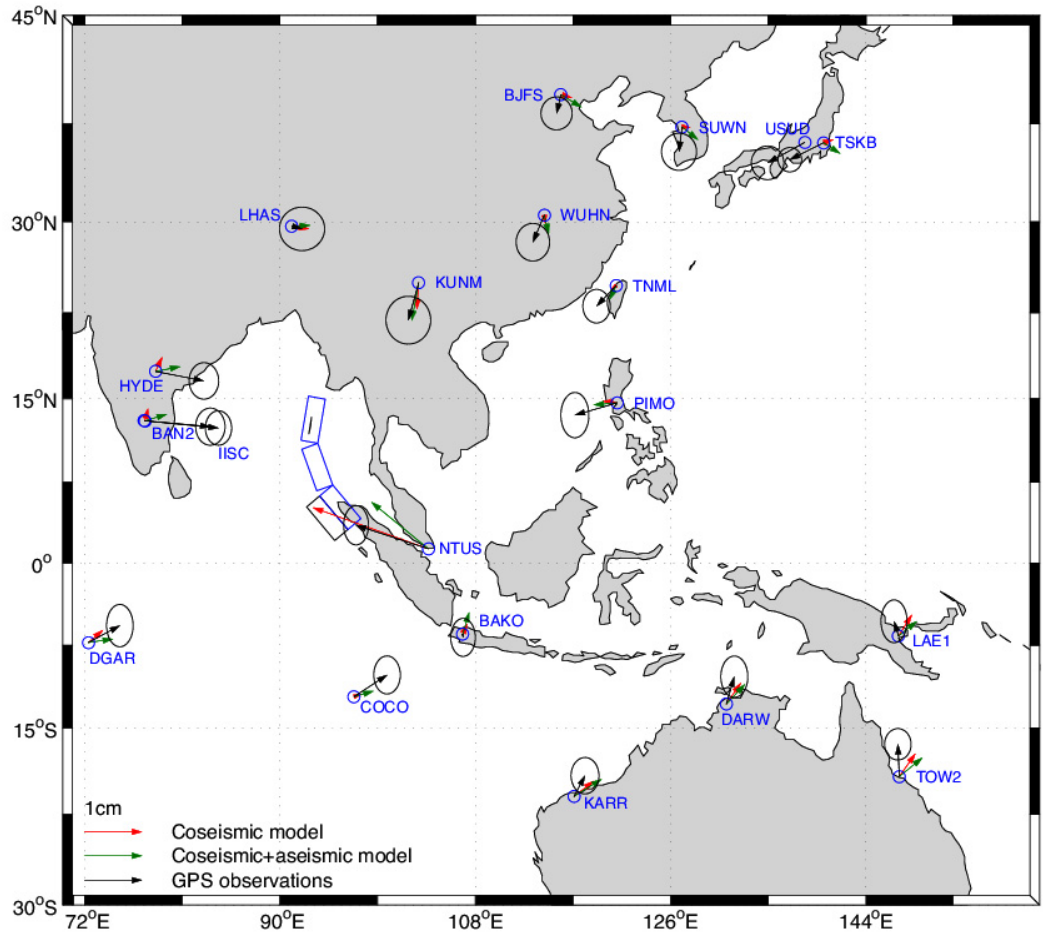

**Figure 1**: Coseismic displacements.

GPS coseismic displacements associated with the Sumatra-Andaman earthquake (black arrows). The largest horizontal displacement of (16.8 ± 2.2) mm is observed at Singapore (NTUS). Red arrows represent the modeled coseismic deformation (4) using source parameters provided by USGS; green arrows represent our best model of combined coseismic and aseismic deformations. The black rectangle represents the fault plane associated with the main event; the three blue rectangles show the sources used for aseismic modeling.

# Supporting Online Material

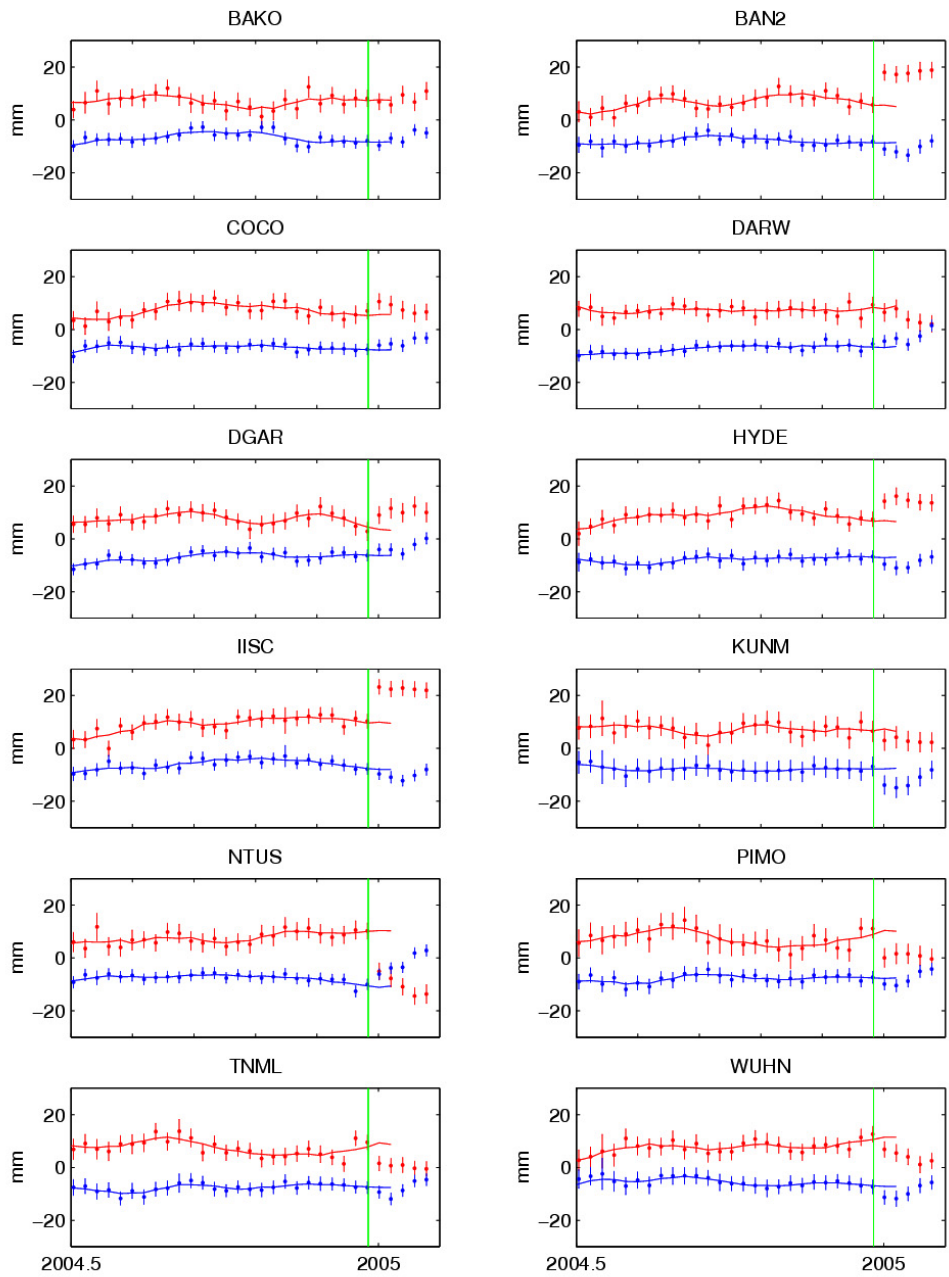

**Figure S1**: GPS Time Series

GPS time series of the East (red) and North (blue) components for selected sites in a large area surrounding the epicenter. The coseismic offset has been computed averaging the positions of two weeks after the event with respect to a running mean filtered series (monthly window). The permanent GPS site at Singapore (NTUS) exhibits the largest horizontal displacement (16.8 ± 2.2) mm in the westward direction (W20°N); a clear displacement of the Indian plate is also evident in three different sites shifting 10 - 14 mm towards the East.

| GPS SITE | Distance from epicentre (km) | 2-D displacement (mm) | Azimuth (deg) | 1-$\alpha$ (%) |
|---|---|---|---|---|
| BAKO | 1,755 | 0.5 ± 2.1 | -108 ± 175 | 69.4 |
| BAN2 | 2,147 | 13.8 ± 2.1 | 97 ± 7 | 99.9 |
| BJFS | 4,597 | 4.3 ± 1.8 | -149 ± 25 | 92.9 |
| COCO | 1,713 | 6.9 ± 2.0 | 58 ± 14 | 96.3 |
| DARW | 4,440 | 4.7 ± 1.6 | -1 ± 29 | 97.7 |
| DGAR | 2,688 | 8.8 ± 2.3 | 68 ± 11 | 99.5 |
| HYDE | 2,336 | 9.8 ± 2.1 | 102 ± 10 | 99.9 |
| IISC | 2,140 | 14.6 ± 2.0 | 94 ± 6 | 99.9 |
| KARR | 3,646 | 3.9 ± 1.8 | 32 ± 29 | 94.1 |
| KUNM | 2,594 | 7.6 ± 2.6 | -162 ± 20 | 99.9 |
| LAE1 | 5,959 | 2.3 ± 2.0 | -41 ± 52 | 90.9 |
| LHAS | 2,959 | 2.3 ± 2.5 | 117 ± 65 | 96.0 |
| NTUS | 1,065 | 16.8 ± 2.2 | -70 ± 5 | 99.9 |
| PIMO | 3,209 | 10.7 ± 2.6 | -102 ± 9 | 99.9 |
| SUWN | 5,054 | 5.1 ± 2.0 | -153 ± 24 | 70.2 |
| TNML | 3,739 | 7.6 ± 1.9 | -117 ± 12 | 99.9 |
| TOW2 | 6,282 | 6.0 ± 1.6 | 24 ± 17 | 97.8 |
| TSKB | 5,952 | 10.4 ± 1.4 | -102 ± 8 | 99.1 |
| USUD | 5,816 | 9.4 ± 1.9 | -110 ± 11 | 99.9 |
| WUHN | 3,701 | 6.2 ± 2.0 | -137 ± 19 | 92.1 |

**Table S1**: Horizontal site displacement estimated from the GPS time series. The weighted mean of the residuals of the two weeks after the mainshock with respect to the running box average (one month window) represent the cumulative displacement after the event with respect to restful plates. The errors are 1-sigma

values (68% confidence region). The last column of the table shows the quantity (1-$\alpha$), where $\alpha$ are the significance values derived from a t-test on the presence of the coseismic displacement. A value of (1-$\alpha$) close to 100% means that the coseismic step is highly significant. It has to be noticed that the significance level should be interpreted with caution since it tests only simple step-wise behaviour in the time series and eventual postseismic effects may also play an important role.